# Frictional damping in radiative electrodynamics and its scaling to macroscopic systems

**D. Das***
(Bhabha Atomic Research Centre
Trombay, Mumbai 400 085.

(e-mail: dasd1951@gmail.com)

Abstract

Radiation force in Abraham-Lorentz-Dirac equation is revisited for possible signature of irreversible action in the dynamics. The analysis shows that the classical electron can 'dissipate' out a certain fraction of field energy that distinguishes itself from the well known Larmor radiation loss. The thermal power loss is shown to follow bi-quadratic acceleration functionality, which is akin to the characteristics of Hawking-Unruh radiation emission from warm surrounding field of a non-inertial observer. Reversibility in nonstationary evolution is possible at the expense of power from concerned external field. By revealing nonlocal mitigation characteristics in nonstationary evolutions, a measure of dissipative relaxation in the radiative electrodynamics is worked out to compare the two distinctly different modes of energy losses. The measure is shown to be applicable uniquely in all scales of externally perturbed systems undergoing nonstationary dynamics, and is used as a common thread to explain frictional contributions of phonons and electrons reported for metals in superconductive phase transitions.

Key words: Abraham-Lorentz-Dirac equation, radiation force, Hawking-Unruh radiation, frictional dissipation.

PACS: 45.20.-d, 03.30.+p

---

*(Presently, superannuated from the official position)

# Frictional damping in radiative electrodynamics and its scaling to macroscopic systems

**(D. Das)**

In macroscopic systems dynamic friction is a common phenomenon where one body sliding over other experiences resistance. The dynamic resistance due to friction is expressed empirically with time asymmetric term proportional to the sliding velocity ($\overline{\mathrm{v}}$). The motion is then described by $m_0\dot{\overline{\mathrm{v}}} = \overline{\mathrm{F}}_{\mathrm{external}} - \kappa\overline{\mathrm{v}}$, where $m_0$, $\overline{\mathrm{F}}_{\mathrm{external}}$ and $\kappa$ are the particle's mass, external force and frictional constant respectively; $\dot{\overline{\mathrm{v}}}$ being observable acceleration. The quantity, $\kappa\mathrm{v}^2/2$ corroborates to frictional power loss. Current approach in understanding the frictional property is based on case specific modeling and numerical simulation analyses. Recently, there is significant progress in measuring the friction property using nano-tip electromechanical probes [1,2,3]. In this study, the frictional power loss is generally described by considering decoherence effect in the sliding operation of one solid surface (probe) over the other (specimen). The driving force in sliding torsionally perturbs the surface phonon modes to result in decoherence, where the transversely driven modes are significantly perturbed and undergo critical damping. Considering the limiting counteraction characteristics of omnipresent nonlocal defense against critical perturbation, it has been possible to generally express the driving force in terms of the basic properties of phonon modes. This helps defining the damping power loss as well as the effective frictional constant.

*Background*

Quantum mechanics has established that dynamic interaction of field and particle in coherent evolution is nonlocally mediated. It further points out that the nonlocally mitigated counteraction to external perturbation has a limit beyond which the coherently evolved system undergoes decoherence. Considering the case of electrodynamics this study shows that at the criticality of decoherence the omnipresent nonlocal field in its incessant effort to reinstate coherent evolution counteracts perturbation and relaxation occurs as a result of stress yield at the supersession of perturbation over the nonlocal defense (details given in the appendix). The mitigation limit in critical counteraction is registered in the radiation recoil effect to perturbation. Considering universality of nonlocal mitigation in decoherence, it is envisaged that the obtained measure of critical reaction to perturbation can lead the way of its generalized description in macroscopic cases such as in the dynamic friction where the coherent phonon oscillations are perturbed leading to dissipation loss under damped oscillations, besides the regular loss of sonic power. As elaborated below, the relaxation features in dynamic friction is quite similar to those in electrodynamics for which the required formulation is comparatively easy because of its rich classical and quantum mechanical bases. Before describing the general formulation, the decoherence characteristics in radiative motion of electrodynamics is briefly mentioned here.

Jerk force influencing upon radiative dynamics of an accelerated charged particle (charge, $q$ and mass, $m_0$) involves the two mutually orthogonal components, $(2q^2/3c^3)\ddot{\bar{\mathrm{v}}} \equiv (2q^2/3c^3)[\ddot{\mathrm{v}}\bar{\mathrm{e}} + \dot{\mathrm{v}}\dot{\bar{\mathrm{e}}}]$; $\ddot{\bar{\mathrm{v}}} \equiv \mathrm{d}(\dot{\mathrm{v}}\bar{\mathrm{e}})/\mathrm{dt}$, $\bar{\mathrm{e}}$ being unit acceleration vector, $\bar{\mathrm{e}} = \dot{\bar{\mathrm{v}}}/\dot{\mathrm{v}}$, and $\dot{\bar{\mathrm{e}}}$ being angular rate arising from torsion of the normal, $\bar{\mathrm{e}}$. The direct component in undergoing displacement results in accelerated growth rate of the kinetic energy ($\mathrm{k} \equiv m_0\mathrm{v}^2/2$) as $(2q^2/3c^3)(\ddot{\mathrm{v}}\bar{\mathrm{e}} \bullet \bar{\mathrm{v}}) =$ $[(2q^2/3m_0c^3)\ddot{\mathrm{k}}$ - $[(2q^2/3c^3)\dot{\mathrm{v}}^2 + \dot{\mathrm{v}}\dot{\bar{\mathrm{e}}}\bullet\bar{\mathrm{v}}]]$. The equation also indicates that the accelerated growth is impeded by two power loss terms: one due to the well recognized Larmor radiation loss while the other from reversed displacement rate ($-\bar{\mathrm{v}}$) of the torsion based jerk component, $(2q^2/3c^3)\dot{\mathrm{v}}\dot{\bar{\mathrm{e}}}$. However, when displacement operation is carried out on direct as well as transverse components of jerk force, the two torsion based power terms cancel out: $(2q^2/3c^3)\ddot{\bar{\mathrm{v}}} \bullet \bar{\mathrm{v}} \equiv (2q^2/3c^3)[(\ddot{\mathrm{k}}/m_0) - \dot{\mathrm{v}}^2]$. This is anyway expected as the transverse force component of jerk contributes to disordered power loss which cannot figure in the displacement based energy conservation analysis.

However, it is to be recognized that the reverse displacement of the charged particle arises from the radiation recoil effect due to the rate of momentum loss $(2q^2\dot{\mathrm{v}}^2/3c^4)\bar{\mathrm{n}}_{\mathrm{rad}}$ of Larmor radiation; $\bar{\mathrm{n}}_{\mathrm{rad}} \equiv (\bar{E} \times \bar{H})/EH$, is the radiation vector, ($\bar{\mathrm{n}}_{\mathrm{rad}}$ being unit radiation vector, $\mathrm{n}_{\mathrm{rad}}^2 = 1$), $\bar{E}$ and $\bar{H}$ (with $E = H$) being transversely oriented electric and magnetic field components in the emitted radiation. Impact due to radiation recoil on mass $m_0$ of the charged particle generates a velocity component given by, $\bar{\mathrm{v}}_{\mathrm{recoil}} = -\dot{\bar{\Pi}}\tau_0/m_0 \equiv -(2q^2\dot{\mathrm{v}}^2\bar{\mathrm{n}}_{\mathrm{rad}}/3c^4)\tau_0/m_0$, where $\tau_0 \equiv (2q^2/3m_0c^3)$ is time involved in recoil-impact, and $\dot{\bar{\Pi}}$ is the rate of momentum loss In Larmor radiation. It is interesting to note that the implied power loss, $\dot{\bar{\Pi}}^2\tau_0/m_0$, is having $\dot{\mathrm{v}}^4$ dependence, which is different in nature as compared to the $\dot{\mathrm{v}}^2$ dependence of Larmor radiation loss. The recoil based power is functionally distinct from the one involved in the power balance equation so far revealed in electrodynamics. The conventional approach of electrodynamics as such cannot sight the possibility for thermal power loss. This could be rooted to the absence of any consideration on dynamic balance in between the two mutually opposing transverse forces arising from jerk and radiation recoil. The unexplored consequence of radiation recoil effect needs further attention as it might have underlying connection with Hawking-Unruh type radiation emission [4,5]. It is important to mention here that in the special case of uniformly accelerated charged particle although jerk has null expectation value ($\langle \ddot{\mathrm{v}} \rangle = 0$), the transverse force component, $(2\mathrm{q}^2/3c^3)\langle \dot{\mathrm{v}}\dot{\bar{\mathrm{e}}} \rangle \equiv \tau_0 \langle m_0\dot{\mathrm{v}}\dot{\bar{\mathrm{e}}} \rangle$, has finite magnitude in the dynamics which is of course nonstationary; finite torsion ($\dot{\bar{\mathrm{e}}}$) results from evolution of unit acceleration vector in the critically perturbed motion with $\mathrm{d}\langle \bar{\mathrm{e}} \rangle/\mathrm{dt} = 0$, $\langle \mathrm{d}\bar{\mathrm{e}}/\mathrm{dt} \rangle \neq 0$. In general, $\dot{\bar{\mathrm{e}}}$ evolves as $\dot{\bar{\mathrm{e}}} = \sum_{\mathrm{k}} \mathrm{c_k}(\text{-i}\omega_{\mathrm{k}})\exp(\text{-i}\omega_{\mathrm{k}}\mathrm{t})$, $\sum_{\mathrm{k}} \mathrm{c_k c_k^*} = 1$. Electrodynamics with constant expectation value of acceleration allows its evolution entailing internal frequency characteristics related to $\tau_0^{-1}$. And the relaxation losses remain unabated in the nonstationary evolution with the observable uniform linear acceleration. Torsion, $\dot{\bar{\mathrm{e}}}$, the pertinent property in disordered relaxation gets established on the polarized vacuum field inherently existing around the point-like charged particle. The spherically polarized field having dynamically relevant radius of $r_0 = 2q^2 / 3m_0c^2$ experiences torque stress of $\bar{F} = \bar{\Gamma}(\dot{e}/c) \equiv (2q^2/3c^3)\dot{v}\dot{\bar{e}}$, $(\bar{\Gamma} \equiv [r_o\bar{n}_r \times m_0\dot{\bar{v}}]$, $\bar{n}_r$ being unit radial vector) in the accelerated dynamics.

As elaborated in the appendix, the dynamic evolution aspect of transverse jerk component is incessantly mitigated by the omnipresent nonlocal field and the mitigation characteristics at critically perturbed dynamics holds the key for the regular (Larmor) and disordered relaxation losses. While the recoil from regular relaxation signifies the manifested nonlocal reaction effect to critical torsional perturbation, the disordered one endorses spontaneity in decoherence. These aspects of nonlocal mitigation characteristics could be revealed by reformulating the electrodynamics purely from classical framework but with the additional measure of optimization of radiation loss/gain in a dynamic course, which is missing in the conventional analysis. When the analysis is based solely on energy-momentum conservation in the radiative interaction of field and (charged) particle, the problem is not closely defined because of no measure on the power in or out of the interaction volume. In absence of the measure, power expense of external source in a dynamic passage is not optimized and left open. The incomplete definition leads to issues like acceleration runaway or generally, in acausality. Classical formalism being blind to nonlocal interplay of nature, restricts its analysis for the field-particle interaction at the radiating boundary (event horizon) at a finite radial distance ($r >> r_0$) from the accelerated charge. As consequence, the analysis could not sight the nonlocal criterion for disordered relaxation. Quantum field theory does recognize the disordered feature using accelerated frame commoving with the charge. According to the commoving observer the surrounding vacuum field of the accelerated charge is thermally populated over ground and excited states; population corroborating to the thermal state predicted by Hawking and Unruh. QFT has not attempted to reveal criticality aspect of the nonlocal mediation in field-particle interaction.

As worked out in the appendix, radiative relaxation occurs in the presence of critical perturbation by transverse force, where nonlocal field attains its limiting defense failing to guard the coherent dynamic state. Relaxation in fact registers the resulting loss of coherence. The criticality is expressible as $\langle \dot{\bar{e}} \rangle_{cr} = \langle \dot{v}/c \rangle_{cr} \bar{n}_{rad}$, $(\bar{n}_{rad} = \lim_{w/c \to 1} \bar{n}_w$, $\bar{n}_w$ being phase velociy vector). It gives a measure of critical torsion that is intrinsically involved in radiation reaction. The manifested torsion leads to thermal relaxation loss in electrodynamics, which occurs along with the regular (Larmor) relaxation at the expense of power from external supply. Thermal relaxation is the result of critically damped oscillations of internal modes of the polarized sphere associated with accelerated charge as the modes are driven by the transverse force. Unlike the regular relaxation, the thermal effect occurs without any work signature. Using standard analytical results of critically damped oscillation [6] driven by a transverse force $\bar{F}$ one finds that energy relaxation occurs at the rate of $(2\xi\mu_0\omega_0)^{-1}F^2\zeta = \dot{E}_{relax}(\text{say})$; $\mu_0$ and $\omega_0$ are respectively the oscillator's effective mass and frequency, $\xi$ the damping parameter which is unity in this case of driven oscillation, and $\zeta = [(2\Delta\omega/\omega_0)^2 + 1]^{-1}$ is the off-resonance coupling efficiency of the driving force, ($\Delta\omega \equiv \omega_0 - \omega_{rad}$ being deviation of the driver's frequency from $\omega_0$;

$\zeta \sim 1$ for near resonance cases ($\Delta\omega << \omega_0$), and $\zeta \sim 1/5$ for far off resonance cases ($\Delta\omega \sim \omega_0$). For electrodynamics, since the rate of change of radiation momentum gives the measure of transverse force one writes $\bar{F} = (2q^2/3c^4)\dot{v}^2\bar{n}_{rad} \equiv \dot{\bar{\Pi}}$, the thermal power loss $\dot{E}_{relax}$ can be rewritten as $\dot{E}_{relax} = [(\dot{\bar{\Pi}} \bullet \delta\bar{\Pi}/m_0)/\pi]\zeta$, where, $\delta\bar{\Pi}/m_0 = \dot{\bar{\Pi}}\tau_0/m_0 \equiv \bar{v}_{recoil}$ is the magnitude of recoil displacement rate, and $\dot{\bar{\Pi}} = (\dot{E}_{Larmor}/c)\bar{n}_{rad} \equiv (m_0 c/\tau_0)(\dot{v}/\dot{v}_{max})^2\bar{n}_{rad}$. It can be seen that the thermal power in torsionally driven damping is expressible as $\dot{E}_{relax} = \zeta[S_0\alpha\hbar\dot{v}^4/6\pi^2c^6]$, where $\alpha \equiv q^2/c\hbar \sim 1/137$ is the fine structure constant, and $S_0\,(=4\pi c^2\tau_0^2)$ is the surface area of the dynamically significant sphere of radius, $c\tau_0$. $\dot{E}_{relax}$ has the bi-quadratic functionality in acceleration indeed. Considering that the relaxation results in thermal radiation emission uniformly over the entire $4\pi$ solid angle of the polarized sphere, the power density expression can be rearranged as $(\dot{E}_{relax}/S_0) = [\zeta(160\alpha)][\sigma(\hbar\dot{v}/2\pi c\kappa_B)^4]$, where $\sigma = \pi^2\kappa_B^4/60\hbar^3c^2$, is the Stefan-Boltzmann constant. The power density corroborates to black body emission temperatre of $T_{rad}(K) = (160\alpha\zeta)^{1/4}[\hbar\dot{v}/2\pi c\kappa_B] \approx 1.04\zeta^{1/4}[\hbar\dot{v}/2\pi c\kappa_B]$, where, $\kappa_B$ is Boltzmann constant. $\zeta$ assumes unity at the highest possible nonlocal defense, which is at the critical acceleration of $\dot{v}_{max}$. With lowering of acceleration, $\zeta$ reduces and converges to the lowest limit of 1/5, $(1/5 \le \zeta \le 1)$. Nonetheless, $T_{rad}$ has semblance to Hawking-Unruh radiation temperature [4,5].

Now, the general form of transverse force $\bar{F}$ can be obtained by considering the oscillation characteristics of the relevant internal modes. As for electrodynamics the dynamically relevant time, $\tau_0 = r_0/c$, corroborates to the internal oscillation frequency as $\omega_0 = 2\pi/4\tau_0$, where $4\tau_0$ expresses time period of one full oscillation that covers sweeping across the diametric stretch of the polarized sphere. As described already, the dynamic evolution involves the transverse force, $\bar{F} = \bar{\Gamma}\dot{e}/c$, $\bar{\Gamma} \equiv [r_o\bar{n}_r \times m_0\bar{v}]\dot{e}$. $\bar{F}$ can be rewritten now using the internal oscillation frequency $(\omega_0)$ as

$$\bar{F} = (\pi/2)(m_0/\omega_0)\dot{v}\dot{\bar{e}}_{cr} \qquad (1).$$

The force expression (1) now involves the characteristic mass, internal mode frequency and the dynamic feature of torsionally stressed polarized sphere. As described below, the expression helps formulating the transverse force in torsionally perturbed phonon modes involved in frictional phenomenon.

*Basic formulation of damping power loss in dynamic friction*

Noting universality of incessantly mitigating nature of nonlocal field to optimize energy losses in critically perturbed dynamics, the relaxation loss in frictional phenomenon is formulated by considering decoherence characteristics under the critical mitigation. Regular relaxation in frictional dynamics across a pair of interacting surfaces occurs through manifestation of sonic energy. In addition to the power loss through sonic propagation there is finite thermal relaxation through the driven damped oscillations of the surface phonon modes. To formulate the frictional damping property in the sliding course of one solid over the surface of another one, it is necessary to define the variables involved in defining decoherence associated with the oscillating phonon modes. Perturbations in the sliding process involve direct driving of the surface phonon modes and also, indirectly through electron plasma oscillation of conduction electrons of the surfaces concerned; the latter one is significant in the cases of electronic conductors like metals/alloys. Driving forces for the direct and indirect ways of mode perturbations are to be defined by considering the general form of transverse force in expression(1) involved in decoherence. Thus using (1), the damping relaxation can be expressed: $\dot{\mathrm{E}}_{\mathrm{relax}}=(2\mu_0\varpi_0)^{-1}F^2\zeta$, where, $\overline{F}=(\pi/2)(\mu_0/\omega_0)\left\langle \dot{\mathrm{v}}\overline{\dot{\mathrm{e}}}_{cr}\right\rangle$, $\mu_0$ and $\varpi_0$ respectively being the reduced mass and frequency of the mode concerned. The acceleration ($\dot{\mathrm{v}}$), and the torsion rate ($\dot{\mathrm{e}}_{cr}$) of phonon modes as involved in $F$ are to be specifically defined for the perturbations. In frictional measurements, the off-resonant coupling efficiency ($\zeta$) of torsional perturbation in surface sliding of solid objects falls under the far off resonance cases, $\zeta \sim 1/5$. This is because the mode oscillation involves significantly high frequency ($\varpi_0$) as compared to those associated with perturbation in the surface sliding operation. Thus, $\dot{\mathrm{E}}_{\mathrm{relax}}=(10\mu_0\varpi_0)^{-1}F^2$.

In the sliding course of friction monitoring probe on the surface of a solid specimen, driving force imparts torsional oscillations of phonon modes of the contact area (in x-y plane, say). In surface oscillation, two orthogonal modes respectively along x and y directions of its $i^{th}$ lattice point are involved according to their coupling angles with the probe movement. Oscillation of surface mode along z direction being not as active as the two other modes of the half space, its effect is generally not considered. But in the description of transversely driven oscillations, it cannot be neglected and it is to be separately considered. The sliding direction making $\vartheta_x$ and $\vartheta_y$ angles with the respective modes leads to torsional oscillations of the two modes in x-y plane; $\vartheta_x+\vartheta_y=\pi/2$. With sliding velocity of $\overline{\mathrm{V}}$, the overall torsional displacement rate of the x and y modes is expressible as $\overline{\dot{\mathrm{e}}}_{cr,i}=(\mathrm{V}/\mathrm{Q}_i)[\hat{\mathrm{i}}_x\cos(\vartheta_x)+\hat{\mathrm{i}}_y\cos(\vartheta_y)]$, ($\overline{\mathrm{i}}_z=\overline{\mathrm{i}}_x\times\overline{\mathrm{i}}_y$; $\mathrm{i}_x^2=\mathrm{i}_y^2=\mathrm{i}_z^2=1$), where $\mathrm{Q}_i$ represents oscillation amplitude ($\overline{\mathrm{i}}_z$ being unit vector representing the x-y plane). Therefore, $\dot{\mathrm{e}}_{cr,\mathrm{i}}^2=(\mathrm{V}/\mathrm{Q}_i)^2$. The above analysis suggests that effectively one of the two surface modes (x- and y-mode) contribute to torsional oscillation. Superscript '$cr$' indicates criticality in torsional oscillations of the phonon modes in the surface sliding process, where the acceleration magnitude involved in the harmonic oscillators is represented by $\dot{\mathrm{v}}\equiv\ddot{\mathrm{Q}}=-\varpi^2 Q$. Thus, the product $(\dot{\mathrm{v}}\overline{\dot{\mathrm{e}}}_{cr})$ has the magnitude of $\varpi^2\mathrm{V}$. Modes are considered as degenerate with fundamental frequency, $\varpi$, and all lattice points are treated to be equivalent. Recalling the expression of thermal loss in damped oscillation which is critically driven by the force $\overline{F}=(\pi/2)(\mu_0/\varpi)\left\langle \dot{\mathrm{v}}\overline{\dot{\mathrm{e}}}_{cr}\right\rangle$ one writes the relaxation rate in an active mode as $\dot{\mathrm{E}}=(10\mu_0\varpi)^{-1}F^2=(\pi^2/40)\mu_0V^2\varpi$. As indicated, $\dot{\mathrm{E}}$ effectively represents thermal power loss from x and y modes of a lattice point. Besides this contribution of the surface modes, sliding action on contact plane of a specimen of finite thickness can torsionally perturb the mode along z direction as indicated already. The damping loss to this additional mode need not double the value expressed above. For this presentation, the half-space contribution from z-mode will be considered later as 50% of $\dot{\mathrm{E}}$ as an approximation.

Unlike the motion of an accelerated charge, the thermal relaxation in sliding motion of macroscopic objects can be significant even at nominal sliding speed ($\mathrm{V}$). This is due to the fact that oscillating phonon modes of the two surfaces in dynamic contact are huge in numbers. Assuming the simpler case of equivalent lattice points, each point contributing one x-y mode, the rate of power loss per unit area over a sliding surface is generally expressed by considering damping loss over the lattice points having surface density $N_s$ as $N_s\dot{E}(\varpi)$, $\dot{E}$ being damping power loss from a lattice point. being Considering the reduced mass $\mu_0$ of a lattice point, and coupling efficiency $\varepsilon$ of the two surfaces in contact in the sliding process one writes, $\dot{\mathrm{E}} = [(\pi^2/40)\varepsilon\mu_0\varpi \mathrm{V}^2]$. Surface roughness together with external load applied normal to the surface decides the coupling. Torsional displacement rate is very much dependent on $\varepsilon$. Stiction and wear are common problems in nano-tip electromechanical probes used in the friction measurement [2]. Considering that each of the modes can vibrate with a frequency distribution from zero to a maximum value, that is, $0 \le \varpi \le \varpi_m$, the above expression of damping loss per unit area needs to be modified by using phonon density of states on surface with frequency of $\varpi$ as $g(\varpi) = (1/\pi)\varpi/\mathrm{v}_s^2$, ($\mathrm{v}_s$ being the speed of stress wave) [7]. Since out of the two distinguishable surfaces modes (longitudinal and transverse), one is effectively perturbed by torsionally driven force, one evaluates the maximum value of actively involved mode frequency, $\varpi_m$, by normalizing the total surface modes per unit area as, $\int_0^{\varpi_m} g(\varpi)d\varpi = N_s$. On integration of RHS, the normalization leads to $\varpi_m = \sqrt{2}\pi^{1/2}\mathrm{v}_s N_s^{1/2}$.

The power loss per unit area is thus expressible as $\dot{E}^{\text{friction}}_{\text{unit area}} = \int_0^{\varpi_m} [(\pi^2/40)\varepsilon^2\mu_0\varpi \mathrm{V}^2]g(\varpi)d\varpi$, where the reduced mass ($\mu_0$) and surface coupling efficiency ($\varepsilon$) are taken to be identical irrespective of the active modes. The integral expression on substitution of $g(\varpi)$ leads to, $\dot{E}^{\text{friction}}_{\text{unit area}} = (\pi/120)\mu_0(\varepsilon \mathrm{V}/\mathrm{v}_s)^2\varpi_m^3$, $\varepsilon$ being the coupling efficiency of the sliding surface, which is the measurement probe of friction property. Using the $\varpi_m$ value in this result one gets the power loss as $\dot{E}^{\text{friction}}_{\text{unit area}} = 0.412\mu_0(\varepsilon \mathrm{V})^2\mathrm{v}_s N_s^{3/2}$. The frictional power loss per unit area can be used now to evaluate the average damping constant in the oscillation of a mode as $\kappa \approx 0.82[\mu_0\varepsilon^2\mathrm{v}_s N_s^{1/2}]$, where the sliding speed of probe is taken as 1cm $s^{-1}$. As discussed an additional 50% over this result is to be considered because of contribution from z-mode in the half-space (sample thickness). Thus, the damping const per lattice point is $\kappa_{\text{lattice point}} \approx 1.23[\mu\varepsilon^2\mathrm{v}_s N_s^{1/2}]$ Thus, for Nb lattice (bulk density, 8.57 $gcm^{-3}$, and atomic weight, 92.9064 a.m.u.), the $N_s$ value is $1.4759\times10^{15}$ $cm^{-2}$, and sound speed is 348000 $cms^{-1}$; the frictional coefficient in phonon mode in the Nb lattice works out to be $\kappa^{Nb}_{\text{lattice point}} \sim 2.5\times10^{-12}\varepsilon^2$ kgs$^{-1}$, $\varepsilon \le 1$. The measured value, if not corrected for the surface-coupling efficiency of measuring probe, will be lower than $2.5\times10^{-12}$ kgs$^{-1}$. For silicon lattice ($\tilde{\mu} \approx 28.0855$ a.m.u., $v_s$ = 843310 cm s$^{-1}$, and $N_s$ =1.35x10$^{15}$ cm$^{-2}$) which is generally used as a probe in the measurement of the friction, the friction constant can be estimated. The frictional coefficient per phonon mode in the Si lattice works out as $\kappa^{Si}_{\text{lattice point}} \sim 1.8\times10^{-12}\varepsilon^2$ kgs$^{-1}$.

In metallic lattice, there are additional components of dissipative losses in the shear action of sliding the friction monitoring probe over specimen surface. The shear force perturbs the oscillating plasma of conduction electron, which in turn can torsionally perturb active modes of positively charged ion cores representing the lattice points. The charged cores undergo oscillations in the plasma field of conduction electrons to be described as $\ddot{\bar{Q}}_{c-plasma} = -[4\pi n_q z q^2/\mu_0]\bar{Q}_{c-plasma} = -[z\omega_p^2(m_0/\mu_0)]\bar{Q}_{c-plasma}$, where $n_q$, $zq$, $\mu_0$, $m_0$, and $\omega_p$ are respectively the number density of conduction electrons, core charge, reduced mass of the core, electron mass, and plasma frequency, $\omega_p = (4\pi n_q q^2/m_0)^{1/2}$. Noting that each core is undergoing harmonic oscillation with the unique frequency of $\omega_{c-plasma} = [z(m_0/\mu_0)]^{1/2}\omega_p$, the action of a shear force in orthogonal direction to the linear oscillations will lead to the frictional dissipation per lattice core atom as, $\dot{E}_{\text{c-}plasma} = 1.5[(\pi^2/40)\mu_0\omega_{c-plasma}(\varepsilon'\text{V})^2]$, $\varepsilon'$ being the coupling efficiency of the sliding surface with the friction monitoring probe. In $\dot{E}_{\text{c-}plasma}$, the factor 1.5 accounts for the number of torsionally active modes per core atom; it is a approximation. Thus, the damping const per core atom is given by $\kappa_{\text{c-}plasma} = 3(\pi^2/40)\mu_0\omega_{c-plasma}\varepsilon'^2$. For the case of Nb metal lattice $n_q$ = 4.33x10$^{22}$ cm$^{-3}$ and $z = 0.79 \pm 0.03$ (at low temperatures) [8], and one thus finds the value of $\omega_p$ as $\omega_p = 1.17\times10^{16}$ s$^{-1}$ and $\omega_{c-plasma} = 2.52\times10^{13}$ s$^{-1}$. Thus, for Nb metal lattice the value of $\dot{E}_{c\text{-}plasma}$ corroborates to frictional coefficient of $\kappa^{Nb}_{c-plasma} \sim 3.0\times10^{-12}\varepsilon'^2$ kgs$^{-1}$. The frictional contribution from electron plasma oscillation alone is significantly low (~8.77x10$^{-15}$ kgs$^{-1}$) and is not considered. The analysis on the whole provides estimates of frictional coefficients using basic solid state properties of matter. Following gives comparison of a meticulously measured data with the results of this analysis.

The estimated value of frictional constant due to electron plasma induced core oscillation in Nb surface, namely, $\kappa^{c-plasma}_{Nb} \sim 3.0\times10^{-12}\varepsilon'^2$ kgs$^{-1}$, $\varepsilon' \le 1$, can be compared with the recently reported value of electronic contribution in Nb lattice [3]. In the superconducting phase transition, the observed decrease in the friction property by about 3.7x10$^{-12}$ kgs$^{-1}$ is quite comparable with the estimated value. Following the transition, the noted residual value of the coefficient of about 1.75x10$^{-12}$ kgs$^{-1}$ is also quite close to the present estimate of the phonon contribution in Nb lattice as $\kappa^{phonon}_{Nb} \sim 2.5\times10^{-12}\varepsilon^2$ kgs$^{-1}$. The disparity is evidently resulting from low surface to probe coupling value, ($\varepsilon \sim 0.7$).

*Appendix: Electrodynamics –in abridged version of classical and quantum mechanics*

In the accelerated motion of a charged particle in external electromagnetic field, the self field loss that occurs at the radiation boundary ($>>$ classical radius of the charge) is the net result of critical negotiation in nonlocally mediated field-particle interaction leading to the optimum relaxation as radiation. This negotiation aspect is not considered in the conventional analysis, because of which the causally interconnected radiation relaxation events remain incompletely specified in the classical description. The incompleteness is reflected by the well marked presence of unwanted acausal consequence like acceleration runaway under withdrawal of external power source abruptly. The nonlocal mitigation involved in optimizing radiation relaxations remaining unaccounted, the power loss ultimately borne by the external field is left open without a measure. Radiation related force designed solely on the basis of energy-momentum conservation falls short of the measure required. The presented analysis proves that adoption of the additional measure of optimizing radiation loss (or, gain) in the dynamic course can describe electrodynamics in a modified form over the well known description by Abraham-Lorentz-Dirac (ALD) equation. The modification of radiation related force involves a nonlocally mediated feature governing relaxation loss in the nonstationary dynamics under external field. Radiation loss (gain) could be generally optimized in the dynamic course variationally defined in between arbitrarily selected pair of space-like surfaces with their time-like separation, only if the course is characteristically endowed with features of delocalized evolution instead of the classically defined course of a world path (line). Using the classical Lagrangian formalism, the delocalized course involving minimum radiation loss/gain in the dynamic passage could be described by linear combination of a set of infinitesimally differed stationary action paths $\{x_p^{\alpha}(\tau)\}$ with nonlocally governed optimal displacement. This is so as the minimum radiation loss (relaxation loss) within a dynamic course can be accomplished under the criterion of optimized path displacement, which inherently involves entanglement of the variational elements, $\delta x^{\alpha} \Leftrightarrow \{\delta x_p^{\alpha}\}$, with instant 4-accelerations of the infinitesimally differed paths at any instant as

$$\delta\int_{\tau_1}^{\tau_2}\sum_p c_p^2 (dx_p^{\alpha} dx_p^{\beta} g_{\alpha\beta})^{1/2} \equiv \sum_p c_p^2 [(g_{\alpha\beta} v_p^{\alpha} \delta x_p^{\beta})\Big|_{\tau_1}^{\tau_2} - \int_{\tau_1}^{\tau_2} (g_{\alpha\beta} \dot{v}_p^{\alpha} \delta x_p^{\beta}) d\tau] = 0 \qquad \text{(A1).}$$

In the displacement optimization equation (A1), the integrated term as defined by the scalars, $\sum_p c_p^2 (v_{\beta,p} \delta x_p^{\beta})$ at $\tau = \tau_1$ and $\tau = \tau_2$, are in fact variations of proper time in the respective cases, which are anyway null by definition of the two time-like boundaries. The remaining last term involves the entanglement of 4-accelerations with the nonzero variation elements. (For the flat space, entanglement is satisfied by a classical geodesic path expressible as $\dot{v}_p^{\alpha} = 0$; geodesic course supports free motion of particles). Thus for general dynamic courses, the measure of minimized relaxation loss implied by eq.(A1), suggests that nonlocal mediation canonically governs the weight factors $\{c_p^2\}$ of the linearly combined paths such that the dynamically relevant 4-acceleration, $\dot{v}^{\alpha}(\tau) \equiv \sum_p c_p \dot{v}_p^{\alpha}(\tau)$, ($\dot{v}^{\alpha} \Leftrightarrow \{\dot{v}_p^{\alpha}\}$, $\sum_p c_p^2 = 1$) corroborates to the 4-orthogoal connection with the variation, $\delta x^{\alpha} \Leftrightarrow \{\delta x_p^{\alpha}\}$, $\delta x^{\alpha} \equiv \sum_p c'_p \delta x_p^{\alpha}$ as $\dot{v}_{\alpha}\delta x^{\alpha} = 0$, ($c_p c_{p'} = \delta_{pp'}$, $\delta_{pp'}$ being Kronecker delta). Replacing the set of variational elements with the corresponding set of space-like unit 4-vectors, $i^{\alpha} \Leftrightarrow \{i_p^{\alpha}\}$, $i^{\alpha} \equiv \sum_p c'_p i_p^{\alpha}$, one rewrites the 4-orthogonal connection as $\dot{v}_{\alpha} i^{\alpha} = 0$, ($i^{\alpha} i_{\alpha} = -1$). Since the nonlocal quantities, $i^{\alpha}$ are independent of displacement, one writes $\dot{v}_{\alpha} i^{\alpha} = 0 = \ddot{v}_{\alpha} i^{\alpha}$. Furthermore, like 4-acceleration and 4-jerk one finds that the canonically averaged 4-velocity, $v^{\alpha} \Leftrightarrow \{v_p^{\alpha}\}$, $v^{\alpha}(\tau) \equiv \sum_p c_p v_p^{\alpha}(\tau)$ is 4-orthogonal to a reduced form of the space-like unit 4-vector, $i'^{\alpha} \equiv (i^{\alpha} - v_{\otimes} v^{\alpha})$, ($v_{\otimes} \equiv i^{\beta} v_{\beta}$). In essence, for meeting the minimum relaxation loss, the nonlocal mediation with dynamical quantities are governed by the following set of connectivities:

$$\dot{v}_{\alpha} i^{\alpha} = 0, \quad \ddot{v}_{\alpha} i^{\alpha} = 0, \text{ and } \quad v_{\alpha} i'^{\alpha} = 0, \; (i'^{\alpha} \equiv i^{\alpha} - v_{\otimes} v^{\alpha}) \qquad \text{(A1a).}$$

The orthogonal connection, $i'^{\alpha} v_{\alpha} = 0$, suggests that the conventionally used radiation 4-force, $(2q^2/3c)[\ddot{v}^{\alpha} + \dot{v}^2 v^{\alpha}]$ needs modification by additional term involving the nonlocal mediation feature $i'^{\alpha}$. The modified 4-force will endorse minimum radiation relaxation in the dynamic passage. The nature of nonlocal mediation and dynamic feature of the envisaged radiation 4-force will be explored now.

Considering now the Langrangian formalism, the stationary path action is represented in the following form:

$$\int_{\tau_1}^{\tau_2} [\sum_p c_p^2 (\partial L_p / \partial x_p^{\mu} + d\pi_{\mu,p} / d\tau - f_{\mu,p}) \delta x_p^{\mu}] d\tau - [\sum_p c_p^2 (\pi_{\mu,p} \delta x_p^{\mu})] \Big|_{\tau_1}^{\tau_2} \equiv \delta S = 0,$$

where, $v_p^{\alpha} \equiv \dot{x}_p^{\alpha}$, $\pi_p^{\alpha} \equiv -(\partial L_p / \partial v_p^{\alpha})$, the canonical 4-momentum on path $p$. Overall Lagrangian in the linear combination is $L(x^{\mu}, v^{\mu}) = \sum_p c_p L_p(x_p^{\alpha}, v_p^{\alpha})$, $L \Leftrightarrow \{L_p\}$. The 4-vector, $f_p^{\alpha}$ represents the radiation related 4-force involved in paths $\{p\}$ of the family. Now, with the consideration of supplementary condition, namely, $\sum_p c_p^2 (\pi_{\mu,p} \delta x_p^{\mu})$, at $\tau_1 = \sum_p c_p^2 (\pi_{\mu,p} \delta x_p^{\mu})$, at $\tau_2$, the delocalized dynamic course of the infinite set of minimally displacing paths $\{x_p^{\alpha}(\tau)\}$ of stationary action is describable irrespective of the arbitrarily selected variational displacements, $\delta x^{\alpha}$ and boundaries ($\tau_1, \tau_2$) involved in the formalism as

$$\partial L / \partial x^{\mu} + d\pi_{\mu} / d\tau - f_{\mu} = 0 \qquad \text{(A2),}$$

where, $f^{\alpha} \Leftrightarrow \{f_p^{\alpha}\}$. $f^{\alpha}$ like the conventional 4-forces being space-like as $f^{\alpha} v_{\alpha} = 0$, eq.(A2) will conserve the rest mass of the charged particle in dynamics. Furthermore, noting that

eq.(A2) endorses that the 4-projection, $f_\alpha \delta x^\alpha$ is governed by that of the conventional terms as, $f_\alpha \delta x^\alpha = (\partial L / \partial x^\alpha + d\pi_\alpha / d\tau)\delta x^\alpha$. Since, the conventional terms, $(\partial L / \partial x^\alpha + d\pi_\alpha / d\tau)$ are local quantities having no components along $\delta x^\alpha$, it leads to

$$f_\alpha i^\alpha = 0 \qquad \text{(A2a).}$$

Functionality of $f^\alpha$ requires detailed consideration of the nonlocal mediator's characteristics, which is ingrained within the supplementary condition, which can be rewritten as

$$(\pi_\alpha i^\alpha)_{\tau_1} = (\pi_\alpha i^\alpha)_{\tau_2} = C \text{ (say)} \qquad \text{(A2b).}$$

This equality is valid irrespective of the arbitrarily selected variation boundaries. The parameter $C$ can be evaluated by considering that at the initial boundary, $\tau_1$ the external field asymptotically disappears and the particle assumes free motion. It is to be noted that for free particle motion with constant 4-velocity components, $v^\beta \equiv \sum_p c_p v_p^\beta$, $(v^2 = \sum_p c_p^2 = 1)$, the evolution equation, (eq.(A1a)), $\dot{v}_\alpha i^\alpha = 0$ corroborates to invariant value of $v_\alpha i^\alpha$ irrespective of the nonlocally evolved quantities, $i^\alpha \Leftrightarrow \{i_p^\alpha\}$. For free motion, the canonically averaged 4-velocity remaining unaltered, the invariancy of $v_\alpha i^\alpha$ is accountable only when the nonlocal evolution in the motion is expressed by the 4-orthogonality, $v_\beta i^\beta = v_\beta \delta x^\beta = 0$. Using $w^\beta \equiv p_0[1, \bar{\mathrm{w}} / c]$, $\bar{\mathrm{w}} / c = \bar{\mathrm{i}}/\mathrm{i}^0$, $i^\alpha \equiv [\mathrm{i}^0, \bar{\mathrm{i}}\,]$, ($p_0$ being a multiplier), the evolution can be rewritten as $v_\beta i^\beta = v_\beta w^\beta = 0$, where, $\bar{\mathrm{w}}$ is nonlocal velocity, or, phase velocity that has magnitudes within, $c \le \mathrm{w} \le \infty$. Since for free motion of a particle of mass $m_0$, the canonical 4-momentum is expressed as $\pi^\mu \equiv m_0 c v^\mu$, the mediator's evolution can be equivalently represented by $\pi_\alpha i^\alpha = 0$. Now recalling back the supplementary condition, (A2b), it is evident that the selected free motion of concerned particle with asymptotic disappearance of the external field at the initial boundary $\tau_1$ is mediated by the evolution $\pi_\alpha i^\alpha = 0$. This makes the parameter $C$ characterizing the supplementary condition (1b) as a null in general and this implies that dynamic evolution with minimum radiation loss corroborates to the 4-orthogonality

$$\pi_\alpha i^\alpha = \pi_\alpha w^\alpha = 0 \qquad \text{(A2c).}$$

Eq.(A2c) speaks for an organized (coherent) evolution of the infinitesimally differed path family $\{x_p^\alpha(\tau)\}$ mediated by the nonlocally defined feature $i^\alpha$.

Now, as per equations in (A1a) the mediating feature $i'^\alpha$ is having its own identity independent of the local hyperplane of 4-tangent, 4-normal $e^\alpha$, ($e^\alpha \equiv \dot{v}^\alpha / \sqrt{-\dot{v}^2}$), and with 4-jerk $\ddot{v}^\alpha$, which has 4-binormal connection as $\sqrt{-\dot{v}^2}\,\dot{e}^\mu \equiv \ddot{v}^\mu + e^\mu(e^\alpha \ddot{v}_\alpha)$, wherein the quantity, $\dot{e}^\mu$ corroborates to 4-binormal, $\dot{e}'^\alpha \equiv (\dot{e}^\alpha - \sqrt{-\dot{v}^2}\, v^\alpha)$, $\dot{e}'^\alpha e_\alpha = 0$, $\dot{e}'^\alpha v_\alpha = 0 = e^\alpha v_\alpha$. Besides 4-orthogonalities with the 4-tangent and 4-normal, ($i'^\alpha v_\alpha = 0$, $e^\alpha i'_\alpha = 0$), the feature $i'^\alpha$ is having finite projection on the 4-binormal as $i'^{\alpha}\dot{e}'_{\alpha} = -v_{\otimes}\sqrt{-\dot{v}^2}$ ; nonzero projection on the local feature $\dot{e}'^{\alpha}$ confirms involvement of the nonlocal mediator in the dynamics. It is now worthwhile to note that the ALD 4-force, $(2q^2/3c)[\ddot{v}^{\alpha}+\dot{v}^2 v^{\alpha}]$, $(\dot{v}^2 \equiv \dot{v}^{\beta}\dot{v}_{\beta})$, is described in the hyperplane of 4-normal and 4-binormal as $(2q^2/3c)[\sqrt{-\dot{v}^2}\,\dot{e}'^{\alpha} - e^{\alpha}(\ddot{v}^{\beta}e_{\beta})]$. Considering this and recalling the 4-orthogonal connection, $i'^{\alpha}v_{\alpha}=0$, one can express radiation related 4-force $f^{\alpha}$ as $f^{\alpha}\equiv(2q^2/3c)[\{\sqrt{-\dot{v}^2}\,\dot{e}'^{\alpha}-e^{\alpha}(\ddot{v}^{\beta}e_{\beta})\}+\phi\dot{v}^2 i'^{\alpha}]$. The space-like 4-force is lying in the local hyperplane of the 4-normal, and 4-binormal ($\dot{e}'^{\alpha}$ ; $\dot{e}'^{\alpha}e_{\alpha}=0$). The parameter, $\phi$ gets normalized to $\phi = v_{\otimes}/(1+v_{\otimes}^2)$ by considering equation (A2a), and also, the equalities: $\dot{e}'^{\alpha}i_{\alpha} = \dot{e}'^{\alpha}i'_{\alpha} = -v_{\otimes}\sqrt{-\dot{v}^2}$ and $i'^{\alpha}i_{\alpha}=i'^{\alpha}i'_{\alpha}=-(1+v_{\otimes}^2)$. Thus, electrodynamics that follows from eq.(1) is expressible as

$$m_0 c\dot{v}_{\mu} - (q/c)F_{\mu\nu}v^{\nu} - (2q^2/3c)[\{\sqrt{-\dot{v}^2}\,\dot{e}'^{\alpha} - e^{\alpha}(\ddot{v}^{\beta}e_{\beta})\}+\phi\dot{v}^2 i'_{\mu}]=0 \qquad \text{(A2d).}$$

In the delocalized representation of electrodynamics eq.(A2d), the dynamic variables involved are to be recalled with their canonically averaged characteristics as was realized over a family of minimally displacing paths of optimum action in a dynamic course. The averaged representation of a quantity $X$ will be generally denoted by $\langle X\rangle$. The canonically averaged representation is necessary to express the nonstationary dynamics which is critically mitigated by the nonlocal field in attaining minimum relaxation loss; mitigating 4-force term being $(2q^2/3c)\phi\dot{v}^2 i'_{\mu}$. The Lorentz 4-force, $(q/c)F_{\mu\nu}v^{\nu}$, being 4-orthogonal to 4-tangent ($v^{\mu}$) and also, to the nonlocal mediator $i'^{\alpha}$, it can be generally described in the hyperplane of 4-normal and 4-binormal. But its canonically averaged projection on 4-binormal, $\langle F_{\mu\nu}v^{\nu}\dot{e}'^{\mu}\rangle$ when expressed in an instant commoving inertial frame as $-\langle \bar{E}\bullet\bar{e}\rangle$ results in null averaged value because of the freely evolving binormal vector $\bar{e}$ under a given external field. In the averaged representation, Lorentz 4-force is oriented solely along 4-normal; $(q/c)\langle(F_{\mu\nu}v^{\nu})\rangle \equiv (q/c)\langle(F_{\alpha\nu}v^{\nu}e^{\alpha})e_{\mu}\rangle$. Thus, by considering this aspect and noting that $f^{\alpha}$ is made of two mutually 4-orthogonal components, $(2q^2/3c)[\{\sqrt{-\dot{v}^2}\,\dot{e}'^{\alpha}\}+\phi\dot{v}^2 i'^{\alpha}]$ and $-(2q^2/3c)(\ddot{v}^{\beta}e_{\beta})e^{\alpha}$, the dynamics can be expressed in the following two components:

$$\langle m_0 c\dot{v}_{\mu} + (2q^2/3c)(\ddot{v}^{\beta}e_{\beta})e_{\mu} - (q/c)(F_{\mu\nu}v^{\nu})\rangle = 0, \qquad \text{(A3),}$$

and,

$$(2q^2/3c)\langle\sqrt{-\dot{v}^2}\,\dot{e}'_{\mu}+\phi\dot{v}^2 i'_{\mu}\rangle = 0 \qquad \text{(A4).}$$

Dynamic variables in the component equation (A3) are oriented along 4-normal. Eq.(A3) represents dynamics in the presence of direct component of jerk 4-force, which in the referred local frame is given by $-(2q^2/3c^3)\langle\ddot{\mathrm{v}}\bar{\mathrm{e}}\rangle$. Its presence results in impediment of the inertial force, $m_0\bar{\ddot{\mathrm{v}}}$ expressible as $\left\langle m_0\bar{\ddot{\mathrm{v}}}[1\text{-}\tau_0 \mathrm{d}\ln(\dot{\mathrm{v}}/\dot{\mathrm{v}}_{\max})/\mathrm{dt}]\right\rangle$, where, $\tau_0 \equiv (2q^2/3m_0c^3) = c/\dot{\mathrm{v}}_{\max}$. Electrodynamics thereby involves modified inertial force, wherein the proper mass of the charged particle gets renormalized to a lower value as $m^* = m_0[1-\tau_0 \mathrm{d}\ln(\dot{\mathrm{v}}/\dot{\mathrm{v}}_{\max})/\mathrm{dt}] \le m_0$. The equality limit for the proper mass holds for the special case of uniformly accelerated dynamics. In this case, one can always find an instantly commoving inertial frame to ensure that there is no change of proper mass in the dynamics ($m^* = m_0$).

Component equation (A4) represents dynamic balance of two 4-forces respectively from torsion based 4-jerk component and the nonlocally mitigating counteraction, $(2q^2/3c)\phi\dot{v}^2 i'_\mu$. The feature of nonlocal counteraction to external perturbation is missing in the conventional description. As mentioned before the torsional perturbation ingrained in jerk evolution is the root of thermal relaxation (similar to Hawking-Unruh effect) in the nonstationary dynamics. In the instant commoving inertial frame, the mitigating force can be further simplified and the dynamic balance (eq.(A4)) can be expressed as $(2q^2/3c^3)[\left\langle\dot{\mathrm{v}}\dot{\bar{\mathrm{e}}}\right\rangle - \left\langle\dot{\mathrm{v}}^2\right\rangle(\bar{\mathrm{n}}_{\mathrm{w}}/\mathrm{w})] = 0$, $(1 \ge c/\mathrm{w} \ge 0)$, where $\bar{\mathrm{n}}_{\mathrm{w}}$ is the unit vector along the phase velocity, $\bar{\mathrm{w}}$. At criticality, the nonlocal reaction to the torsional perturbation attains its highest value as $c/\mathrm{w}$ approaches unity. Thus the critical balance is represented by $\left\langle\dot{\bar{\mathrm{e}}}\right\rangle_{cr} = \left\langle\dot{\mathrm{v}}/\mathrm{c}\right\rangle_{cr}\bar{\mathrm{n}}_{\mathrm{rad}}$, ($\bar{\mathrm{n}}_{\mathrm{rad}} = \lim_{w/c\to 1}\bar{\mathrm{n}}_{\mathrm{w}}$, $\bar{\mathrm{n}}_{\mathrm{rad}}$ being radiation vector). The result gives a measure of critical torsion that is intrinsically involved in radiation reaction. In the changeover of coherent to incoherent evolution occurring at critical perturbation, the manifested torsion leads to dissipative relaxation loss in electrodynamics, which occurs along with the regular (Larmor) relaxation at the expense of power from external supply. Recoil momentum from regular relaxation is now identifiable with the nonlocal reaction, $-(2q^2/3c^3)\left\langle\dot{\mathrm{v}}^2\right\rangle(\lim_{\mathrm{w}\to c}\bar{\mathrm{w}}/\mathrm{w}^2)$. For the special case of uniformly accelerated linear motion of a charge ($\left\langle\ddot{\mathrm{v}}\right\rangle = 0$), the torsion component is inherently present as $\left\langle\bar{\ddot{\mathrm{v}}}\right\rangle \equiv \left\langle\mathrm{d}(\dot{\mathrm{v}}\bar{\mathrm{e}})/\mathrm{dt}\right\rangle = \left\langle\ddot{\mathrm{v}}\right\rangle\bar{\mathrm{e}} + \left\langle\dot{\mathrm{v}}\dot{\bar{\mathrm{e}}}\right\rangle$. Therefore, the relaxation losses remain unabated in the nonstationary evolution with the observable uniform linear acceleration. It is important to note that the presence of relaxations abhors acceleration runaway as an alternative solution to the dynamics under abrupt withdrawal of external field, prohibiting thereby acausality issue.

The delocalized description of dynamics arrived in (A2) could be rationalized after understanding the canonical governance of the nonlocal mediator $i^\alpha$ expressed by the supplementary condition, $\pi_\alpha i^\alpha = \pi_\alpha w^\alpha = 0$. Now, recognizing that the velocity-like 4-vector, $w^\alpha \equiv \lambda[1, \bar{\mathrm{w}}/c]$, $1 \ge c/\mathrm{w} \ge 0$, corroborates to the phase velocity, its 4-components in the reciprocal space coordinates is expressible as $w_k^\alpha \equiv p_0[1, \omega_{\mathrm{k}}/c\bar{\mathrm{k}}]$, where $k^\alpha \equiv [\omega_{\mathrm{k}}/c, \bar{\mathrm{k}}]$, ($\omega_{\mathrm{k}}/c \equiv \mathrm{k}^0$, and $p_0$ is a multiplier). $w_k^\alpha$ being perennially 4-orthogonal to the spectral 4-cordinates as $k^\alpha w_{\alpha,k} = 0$, one finds that the spectrally represented supplementary condition, $\pi_{\mathrm{k}}^\alpha w_{\alpha,\mathrm{k}} = 0$, has the following characteristics: $\pi_{\mathrm{k}}^\alpha$, evolves in the reciprocal space as $\pi_k^\alpha \propto k^\alpha$. Since this spectral connection of the nonlocal mediator stands irrespective of the particle, external field, and their interactions, the proportionality constant is versatile in nature. The proportionality constant is having the unit of action, which can be identified with the Planck's constant, $\hbar$. The canonical 4-momentum thus evolves in quantized form as $\pi_k^\alpha = \hbar k^\alpha$. Noting that in electrodynamic evolution, $\pi_k^\alpha$ can be expressed as $\pi_k^\alpha = m_0 c v_k^\alpha + qA^\alpha / c$, ($A^\alpha$ being 4-potential components of electromagnetic field), it is seen that the scalar representation of canonical 4-momentum evolution is having the form $\sum_k \sum_{k'} a_k a_{k'} [(\hbar k^\alpha - qA^\alpha / c)(\hbar k'_\alpha - qA_\alpha / c) - m_0^2 c^2 v_k^\alpha v_{\alpha,k'}] = 0$, with $a_k a_{k'} = \delta_{kk'}$. Noting that the summation $\sum_k \sum_{k'} a_k a_{k'} (v_k^\alpha v_{\alpha,k'})$ in the evolution equation represents the scalar $(v^\alpha v_\alpha)$, which is unity, the result corroborates to the dispersion property of the nonlocal mediator ($w^\alpha$) as $(\hbar k^\alpha - qA^\alpha / c)^2 = m_0^2 c^2$. Phase velocity and group velocity of the mediator are obtainable from the dispersion equation. Dynamic equation (A2) along with the dispersion relation governs the canonical rule of the delocalized passage of the (scalar) charged particle in the abridged (classico-quantum) version of the arrived electrodynamics. Dispersion characteristics of the scalar evolution under non-relativistic approximation can be similarly represented. Dispersion characteristics for spinor evolution case can also be obtained by involving Dirac matrices in expressing the quantized canonical 4-momentum of the spinor particle.

The canonical rules in the declocalized description of dynamics though arrived purely from the consideration of nonlocal mitigation feature for minimized relaxation characteristics do conform to the well established quantum mechanical prescriptions. (It is worthwhile to mention here that the deductively obtained description of delocalized dynamic has similarity with the well known Feynman's approach[#] of summed over phase histories of all possible paths). For dynamic passage that involves no relaxation loss, the canonical rule of coherent evolution shows null expectation values of acceleration and jerk, and thus ensures the absence of the radiation related force. However, with the loss of coherence under critically high external field, the expectation values of acceleration and jerk are finite to result in radiation relaxations of the ordered and disordered kinds respectively as Larmor radiation loss and thermal radiation loss from the Hawking-Unruh effect.

([#]Feynman R P and Hibbs A R (1965) Quantum Mechanics and Path Integrals, New York: McGraw-Hill)